\documentclass{INTERSPEECH2023}

\usepackage{color,xcolor}
\usepackage{epsfig}
\usepackage{graphicx}
\usepackage{lipsum}
\usepackage{pifont}
\usepackage{array}
\usepackage{booktabs}
\usepackage{colortbl}
\usepackage{multirow}
\usepackage{float}
\usepackage{caption}
\usepackage{adjustbox}
\usepackage{subfigure}
\usepackage{footnote}
\makesavenoteenv{tabular}
\makesavenoteenv{table}

\usepackage{amsmath,amsfonts,amssymb,bm}
\usepackage[super]{nth}

\usepackage{changepage}
\usepackage{extramarks}
\usepackage{lastpage}
\usepackage{setspace}
\usepackage{soul}
\usepackage{xspace}

\usepackage{url}
\usepackage{hyperref}
\hypersetup{colorlinks=True, urlcolor=black}

\usepackage{algorithm, algorithmic}
\usepackage{enumitem}
\usepackage{verbatim}
\usepackage{pifont}
\usepackage[acronyms]{glossaries}
\glsdisablehyper

\newcommand{\eqndot}[1]{Eqn.~(\ref{eqn:#1})}

\newcommand{\figdot}[1]{Fig.~\ref{fig:#1}}
\newcommand{\tbl}[1]{Table~\ref{tab:#1}}

\newcommand{\ignore}[1]{}

\makeatletter
\DeclareRobustCommand\onedot{\futurelet\@let@token\@onedot}
\def\@onedot{\ifx\@let@token.\else.\null\fi\xspace}

\makeatother

\definecolor{MyDarkBlue}{rgb}{0,0.08,1}
\definecolor{MyDarkGreen}{rgb}{0.02,0.6,0.02}
\definecolor{MyDarkRed}{rgb}{0.8,0.02,0.02}
\definecolor{MyDarkOrange}{rgb}{0.40,0.2,0.02}
\definecolor{MyPurple}{RGB}{111,0,255}
\definecolor{MyRed}{rgb}{1.0,0.0,0.0}
\definecolor{MyGold}{rgb}{0.75,0.6,0.12}
\definecolor{MyDarkgray}{rgb}{0.66, 0.66, 0.66}

\usepackage{pgfplots}
\DeclareUnicodeCharacter{2212}{−}
\usepgfplotslibrary{groupplots,dateplot}
\usetikzlibrary{patterns,shapes.arrows}
\pgfplotsset{compat=newest}

\interspeechcameraready
\title{Blank-regularized CTC for Frame Skipping in Neural Transducer}
\name{
Yifan Yang\textsuperscript{*,1},
Xiaoyu Yang\textsuperscript{*,1},
Liyong Guo\textsuperscript{1},
Zengwei Yao\textsuperscript{1}, \\
Wei Kang\textsuperscript{1},
Fangjun Kuang\textsuperscript{1},
Long Lin\textsuperscript{1},
Xie Chen\textsuperscript{\dag,2},
Daniel Povey\textsuperscript{\dag,1}
\thanks{
* stands for equal contribution. \dag~stands for corresponding authors.
}
}
\address{\textsuperscript{1}Xiaomi Corp., Beijing, China \hspace{1em} \textsuperscript{2}Shanghai Jiao Tong University, Shanghai, China}
\email{\{yangyifan7, yangxiaoyu6, guoliyong, yaozengwei, kangwei1, kuangfangjun, linlong, dpovey\}@xiaomi.com, chenxie95@sjtu.edu.cn}

\begin{document}

\maketitle

\begin{abstract}
Neural Transducer and connectionist temporal classification (CTC) are popular end-to-end automatic speech recognition systems. Due to their frame-synchronous design, blank symbols are introduced to address the length mismatch between acoustic frames and output tokens, which might bring redundant computation. Previous studies managed to accelerate the training and inference of neural Transducers by discarding frames based on the blank symbols predicted by a co-trained CTC. However, there is no guarantee that the co-trained CTC can maximize the ratio of blank symbols. This paper proposes two novel regularization methods to explicitly encourage more blanks by constraining the self-loop of non-blank symbols in the CTC. It is interesting to find that the frame reduction ratio of the neural Transducer can approach the theoretical boundary. Experiments on LibriSpeech corpus show that our proposed method accelerates the inference of neural Transducer by 4 times without sacrificing performance.

\end{abstract}
\vspace{-0.25em}
\section{Introduction}
End-to-End (E2E) architectures are gaining more and more attraction in the field of automatic speech recognition (ASR). Several prominent E2E architectures are developed in recent years, such as Connectionist Temporal Classification (CTC)~\cite{CTC}, Attention-based Encoder-Decoder (AED)~\cite{AED}, and neural Transducer~\cite{RNN-T}. Among these three models, CTC and neural Transducer share some common characteristics since they are frame-synchronized systems, where each acoustic frame is mapped to one or more output tokens. In contrast, label-synchronized decoding is adopted in AED, where one valid label token is generated at every step.
Due to its streaming nature and superior performance in a range of tasks, the neural Transducer model receives increasing attention from both academic research and industry application. However, compared to the AED model, the decoding of the neural Transducer is more computationally expensive since it needs to handle the output of each frame in the frame-synchronized decoding.

As the input sequence of acoustic frames is typically much longer than the target label sequence, a special blank symbol corresponding to ``output nothing'' is introduced in the frame-synchronous RNN-T and CTC architectures. During inference, a large proportion of the acoustic frames are classified as blank frames, which can be a waste of computation. For the purpose of accelerating decoding, various studies have examined the identification of blank frames and the influence of discarding blank frames on the decoding results.
Chen et al.~\cite{PSD} investigated the peaky posterior property of a CTC model and found that blank frames contribute little to decoding performance. Similarly, Zhang et al.~\cite{RNN-T_PSD} skipped blank labels to speed up the neural Transducer decoding process. Tian et al.~\cite{NLPRBS} discarded encoder output frames based on the blank probabilities generated by a co-trained CTC to reduce the number of encoder frames going through the joiner only during inference. Similar to \cite{NLPRBS}, Wang et al.~\cite{GoogleBS} applied frame skipping in the middle of the shared encoder during training and inference.

The inference acceleration of the neural Transducer model is achieved by discarding blank frames the co-trained CTC predicted~\cite{NLPRBS, GoogleBS}. If a larger proportion of acoustic frames can be accurately classified as blank frames by the co-trained CTC, the inference speed of the neural Transducer can be further optimized. To this end, we explored various methods to explicitly regularize the blank probabilities the CTC predicted. The CTC branch is encouraged to emit more blank frames by applying a penalty $\lambda$ on consecutively repeated non-blank labels or constraining the maximum number $K$ of consecutively repeated non-blank labels in the CTC topology. We show that the number of non-blank acoustic frames can approximate the target label tokens by adjusting $\lambda$ or $K$. Experiments on the LibriSpeech corpus demonstrate that the neural Transducer with guidance from blank-regularized CTC yields an even lower word-error-rate (WER) than the baseline model without discarding blank frames. Our approaches achieve better trade-offs between WER and inference speed than existing methods.

To summarize, our contributions are three-fold:
\begin{itemize}
    \item We propose two novel regularization methods to explicitly encourage the blank symbols in the co-trained CTC, which can further speed up the inference of neural Transducer.
    \item By applying our proposed strategies, the frame reduction ratio of the neural Transducer could even approach the theoretical boundary.
    \item Experimentally, we achieve an inference speedup of 4 times compared to the standard neural Transducer without sacrificing performance, and a speedup of 1.5 times over a competitive baseline~\cite{GoogleBS} using frame reduction.
\end{itemize}

This work uses the k2~\cite{k2} framework\footnote{https://github.com/k2-fsa/k2} for modifying CTC topology and loss computation. The code is released as part of the open-source project icefall\footnote{https://github.com/k2-fsa/icefall}.
\vspace{-0.25em}
\section{CTC and Neural Transducer}
\subsection{Connectionist Temporal Classification}
CTC~\cite{CTC} is one of the earliest E2E ASR frameworks, which comprises an encoder and a linear decoder.
To address the length mismatch between acoustic frames and output token sequences, the output vocabulary $\mathcal{V}$ is augmented by a blank symbol $\varnothing$ representing no label emission.
Given a sequence of input acoustic features $\boldsymbol{x} = \left(x_{1}, \cdots, x_{T}\right)$ of length $T$, the encoder produces embeddings $\boldsymbol{f} = \left(f_{1}, \cdots, f_T\right)$.
The embeddings are then passed through the CTC decoder to generate $T$ conditionally independent posterior probabilities $p_1,\cdots,p_T$, corresponding to the vocabulary $\mathcal{V} \cup \{\varnothing\}$.
Given a ground truth label sequence, $\boldsymbol{y} = \left(y_{1}, \cdots, y_{U}\right)$ of length $U$, $y_u \in \mathcal{V}$, the CTC objective function is defined as the probability of all possible alignments between $\boldsymbol{x}$ and $\boldsymbol{y}$:

\vspace{-0.3em}
\begin{equation}
\label{eqn:CTC_object_function}
\mathcal{L}(\boldsymbol{y})=\sum_{\boldsymbol{\pi} \in \mathcal{B}^{-1}(\boldsymbol{y})} \log p\left(\boldsymbol{\pi} \mid \boldsymbol{x} \right),
\vspace{-0.2em}
\end{equation}
where $\mathcal{B}(\cdot)$ is a many-to-one mapping that removes repetitive labels and blank labels in an alignment.

With the conditional independence assumption of CTC, the objective function \eqndot{CTC_object_function} can be approximated as:
\vspace{-0.3em}
\begin{equation}
\label{eqn:CTC_object_function2}
\mathcal{L}(\boldsymbol{y}) \approx \sum_{\boldsymbol{\pi} \in \mathcal{B}^{-1}(\boldsymbol{y})} \sum_{t=1}^{T} \log p\left(\pi_{t} \mid \boldsymbol{x}\right)
\vspace{-0.4em}
\end{equation}

Weighted Finite-State Transducer (WFST) topologies can be employed to implement CTC-like algorithms efficiently~\cite{EESEN, NvidiaCTC}. 
A training lattice involves three graphs:
\begin{itemize}
    \item a topology graph ($\mathbf{H}$) that acts as the map $\mathcal{B}(\cdot)$;
    \item a lexicon graph ($\mathbf{L}$) that maps sequences of lexicon units to words;
    \item a dense Finite State Acceptor ($\mathbf{Dense.FSA}$) whose weights represent the acoustic log-probabilities. 
\end{itemize}
The CTC lattice can be obtained in two steps. First, the $\mathbf{H}$ and $\mathbf{L}$ are composed to create the $\mathbf{HL}$ graph, which will convert the predicted token sequences into word sequences. Then, the $\mathbf{HL}$ graph is intersected with $\mathbf{Dense.FSA}$ to produce the CTC lattice, which contains all valid alignments $\mathcal{B}^{-1}(\boldsymbol{y})$.
Instead of using the traditional forward-backward algorithm~\cite{CTC}, we can compute the total score of the CTC lattice with a differentiable dynamic programming method\footnote{Fsa.get\_tot\_scores() in https://github.com/k2-fsa/k2} for optimizing the CTC objective function as shown in \eqndot{CTC_object_function2}.

\vspace{-0.2em}
\subsection{Neural Transducer}
Neural Transducer~\cite{RNN-T} is proposed to address the conditional independence assumption in CTC, where the output probability of $y_u$ is conditioned on all the previous tokens $\boldsymbol{y}_{\leq {u-1}}$. The decoder which functions like a language model is always fed with the previously emitted token. The joint network defines the probability $p(k|t,u)$ of emitting token $k$ at time $t$ after emitting $u-1$ previous tokens by fusing the acoustic embedding and text embedding. Similar to CTC, neural Transducer models also maximize the probability $p(\boldsymbol{y}|\boldsymbol{x})$ (referred to as RNN-T loss) by summing over all possible alignments:
\vspace{-0.2em}
\begin{equation}
\label{eqn:RNNT_object_function}
\mathcal{L}(\boldsymbol{y})=\sum_{\boldsymbol{\pi} \in \mathcal{A}^{-1}(\boldsymbol{y})} \log p\left(\boldsymbol{\pi} \mid \boldsymbol{x} \right),
\vspace{-0.4em}
\end{equation}
where $\mathcal{A}(\cdot)$ is a many-to-one mapping that only removes blank labels in an alignment.

It is worth noting that the blank symbols in CTC and neural Transducer serve very similar functions by separating two succeeding label tokens and aligning the input acoustic frames and output label tokens. There are still minor differences between these two types of blank symbols. Specifically, repeated label tokens are allowed in the CTC alignments and a blank symbol is also necessary to distinguish two consecutive identical label tokens, while in the neural Transducer, each label token is produced only once and the rest of the output tokens are blank symbols in each alignment. It is expected that the blank symbol predictions of these two models are highly relevant and well synchronized. 

\vspace{-0.2em}
\subsection{CTC-guided Neural Transducer}
Motivated by the similar role of blank symbols in CTC and neural Transducer, several previous works attempted to utilize blank symbols from CTC to guide and simplify the inference of neural Transducer systems.
Tian et al.~\cite{NLPRBS} introduced a method to discard encoder output frames based on the blank probabilities generated by the CTC branch. By reducing the number of encoder output frames fed into the joiner, the decoding speed can be largely improved. However, blank skipping is not performed during training. This mismatch between the ways of processing blank frames during training and inference leads to sub-optimal performance.
To achieve a coherent frame-skipping behavior between training and inference, Wang et al.~\cite{GoogleBS} applied frame skipping in the middle of the shared encoder during training based on the blank probability predicted by a co-trained CTC model.
During the forward pass, the CTC branch computes the blank probabilities and frames with a blank probability higher than the predetermined threshold will be excluded from the RNN-T loss computation.
The authors reported a significant speed-up of training and inference with the neural Transducer model, without performance degradation.

\begin{figure}
    \centering
    \includegraphics[scale=0.6]{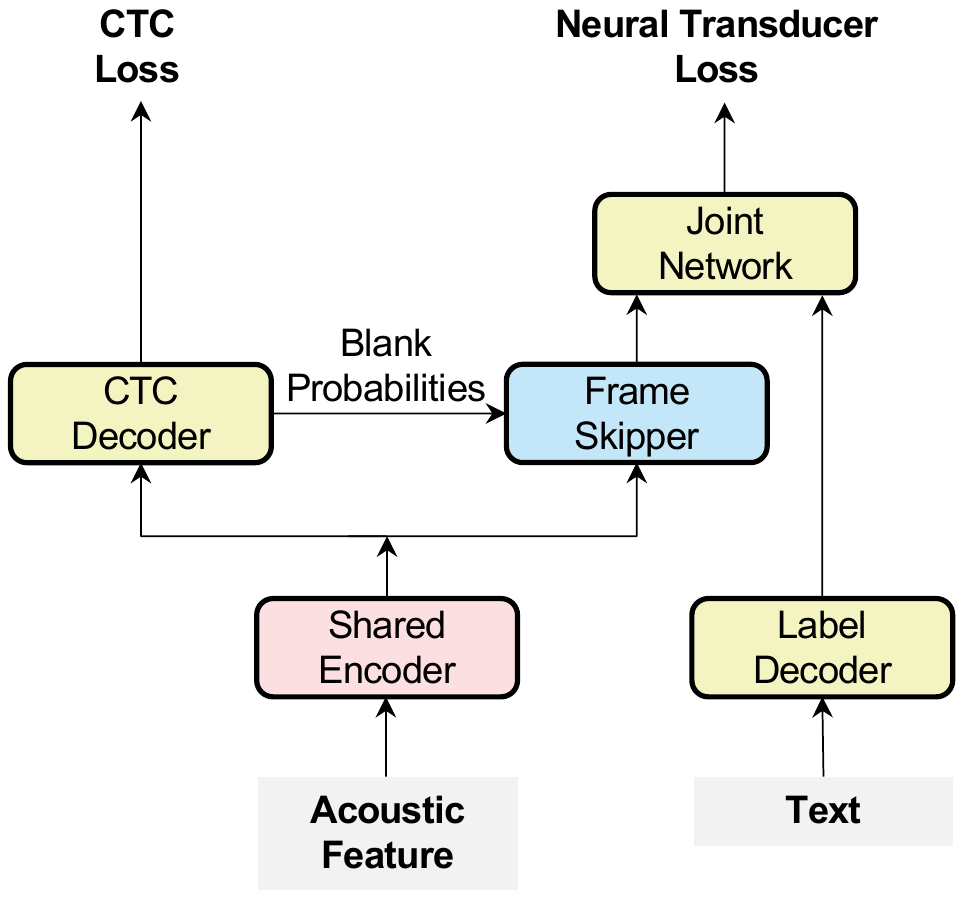}
    \caption{System architecture of neural Transducer with a co-trained CTC branch for blank skipping.}
    \label{fig:system}
    \vspace{-2em}
\end{figure}

\vspace{-0.25em}
\section{Methods}
\begin{figure}[t]
\centering
\subfigure[standard $\mathbf{HL}$] {
    \label{fig:raw}
    \includegraphics[scale=0.35]{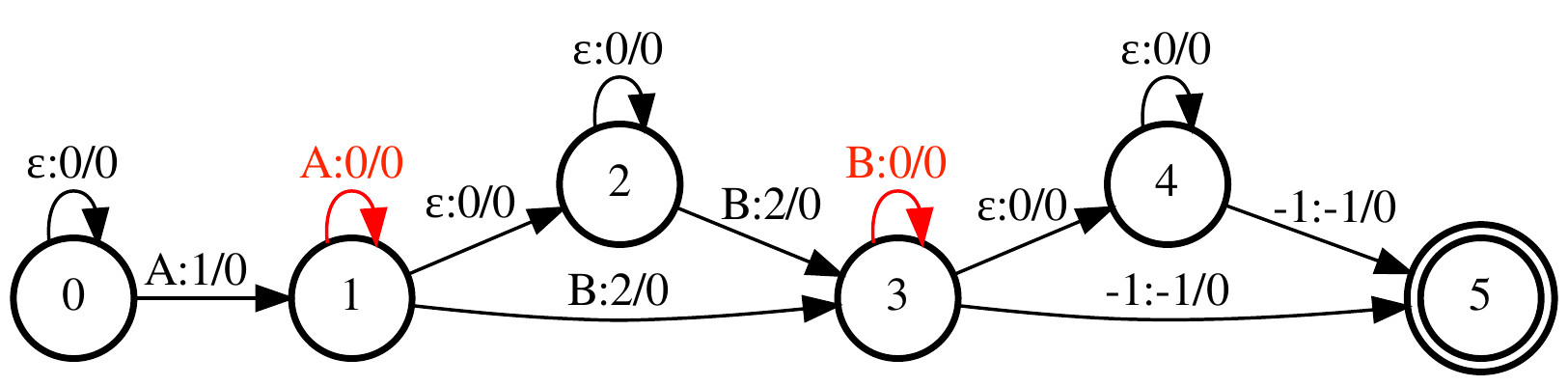}
}
\subfigure[$\mathbf{HL}$ with soft restriction $\lambda$] {
    \label{fig:soft}
    \includegraphics[scale=0.35]{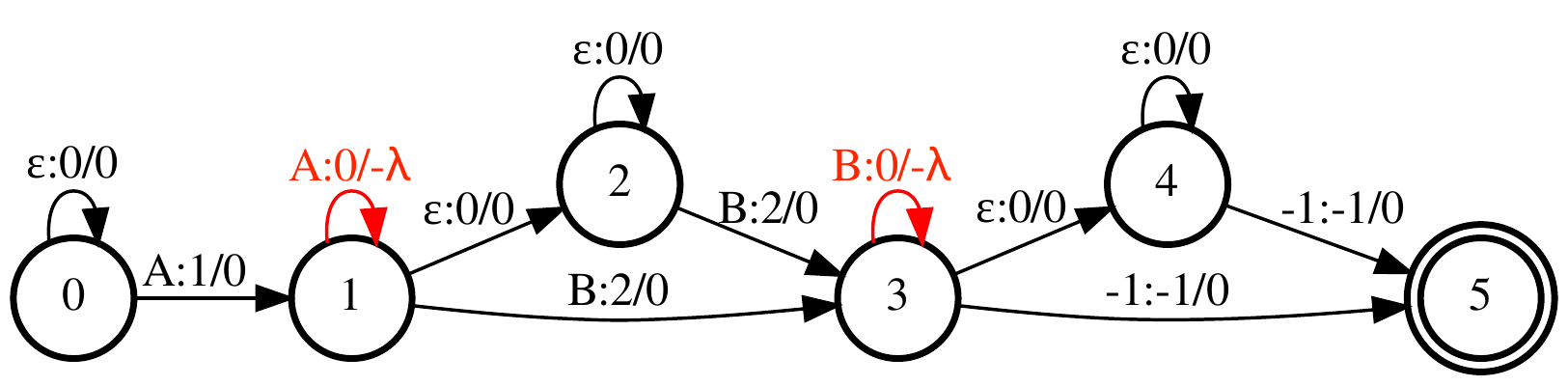}
}
\subfigure[$\mathbf{HL}$ with hard restriction, $K=1$] {
    \label{fig:hard_1}
    \includegraphics[scale=0.35]{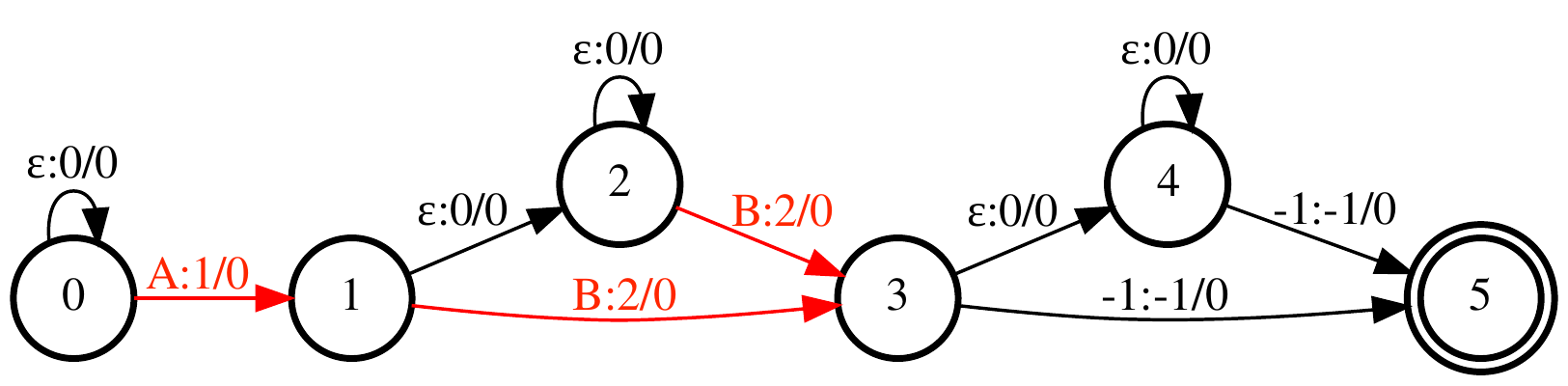}
}
\subfigure[$\mathbf{HL}$ with hard restriction, $K=2$] {
    \label{fig:hard_2}
    \includegraphics[scale=0.35]{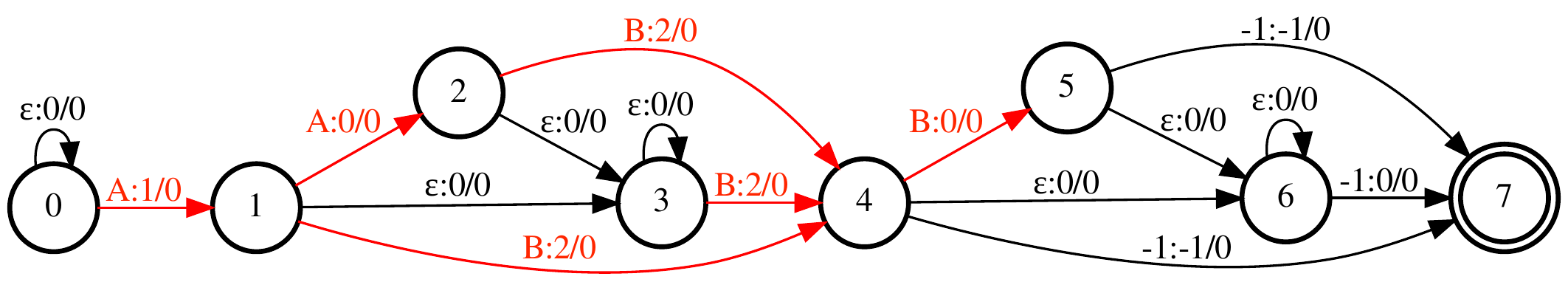}
}
\caption{WFST representations of CTC \textbf{HL} graph outputting the label sequence ``AB'' w/o blank regularizations. An arc with the label ``\textbf{a}:\textbf{b}/\textbf{s}'' means the WFST consumes the input token \textbf{a} and emits the output token \textbf{b} with score \textbf{s}. Arcs entering the final state have ``\textbf{-1}:\textbf{0}/\textbf{0}'' as the label. For soft restriction, a penalty $\lambda$ is applied over the self-loop of all non-blank symbols. For hard restriction, the maximum number $K$ of consecutively repeated non-blank symbols is constrained. Note for $K=1$, successive repeats of non-blank symbols are not allowed.}
\label{HL}
\vspace{-2em}
\end{figure}

Existing methods~\cite{NLPRBS, GoogleBS} improve the inference speed of neural Transducers by skipping blank frames with the help of a co-trained CTC. However, few studies have explored the feasibility of skipping non-blank frames.
According to the definition of $\mathcal{B}(\cdot)$ in Eqn. \ref{eqn:CTC_object_function}, the CTC model not only removes blank symbols but also merges consecutively repeated label tokens. Supposing that the frames emitting repeated non-blank symbols can be treated as blank frames by the CTC branch, more encoder output frames can be discarded in the neural Transducer, which will lead to a further inference speedup.
Inspired by this observation, two strategies are proposed to narrow the spikes of non-blank posteriors in CTC and force the model to output fewer consecutively repeated non-blank tokens.

\subsection{Soft Restriction}
The first proposed strategy employs an additional fixed penalty $\lambda$ (e.g 0.05) to all non-blank self-loops in the standard \textbf{HL} graph of CTC during training, as shown in~\figdot{raw}. Consequently, the alignments containing more consecutively repeated non-blank symbols will receive larger penalties, as illustrated in~\figdot{soft}, and the CTC model tends to encourage alignments with fewer consecutively repeated non-blank symbols. It is worth noting that we only apply this non-blank self-loop penalty $\lambda$ during the training. The proportion of the blank frames from the CTC branch can be controlled by tuning the penalty $\lambda$. This strategy is referred as \textit{soft restriction}.

\subsection{Hard Restriction}
In \textit{soft restriction}, the CTC branch yields a larger proportion of skipped frames (i.e. blank frames) by penalizing CTC alignments with consecutively repeated non-blank symbols. However, the alignments having different numbers of repeated non-blank tokens are still optimized during training as these are valid alignments by definition.
Another strategy named \textit{hard restriction} can be applied to explicitly limit the maximum number of consecutively repeated non-blank symbols $K$ (including the first non-blank symbol) during training. The resulting WFST topologies for $K=1$ and $K=2$ are given in \figdot{hard_1} and \figdot{hard_2} respectively. As can be seen, the alignments with more than $K$ consecutively repeated non-blank symbols are pruned from the corresponding $\mathbf{HL}$ graph.

\subsection{Frame Skip}
During training, the encoder output frames are filtered based on the blank probabilities $p_t^{\varnothing}$ predicted by the co-trained CTC branch. If $p_t^{\varnothing}$ surpasses a pre-defined threshold $\beta$ (e.g 0.85), the $t$-th frame is classified as a blank frame and discarded. from the RNN-T loss computation. The set of discarded encoder output frames can be described as follows:
\begin{align}
    \boldsymbol{f}_{skipped} = \left \{ f_t \mid p_t^{\varnothing} \leq \beta \right \}.
\end{align}
During inference, a trade-off between frame reduction ratio and recognition accuracy can be tuned by adjusting the threshold $\beta'$ for blank skipping.
\vspace{-0.25em}
\section{Experiments}
\subsection{Experimental Setups}
\noindent\textbf{Dataset} The LibriSpeech~\cite{LibriSpeech} corpus is used for experiments, containing 960 hours of transcribed audiobook recordings. Lhotse~\cite{Lhotse} is employed to perform data preparation. Speed perturbation~\cite{SpeedPerturb} with factors of 0.9 and 1.1 is applied to augment the training data. SpecAugment~\cite{SpecAugment} and noise augmentation~\cite{musan} are utilized to improve model robustness in training. The input features are 80-channel filter bank features extracted from 25 ms windows with 10 ms shift. The classification units are 500-class Byte Pair Encoding (BPE)~\cite{bpe} word pieces. 

\noindent\textbf{System Architecture} The co-trained CTC\&Transducer model as shown in \figdot{system} is adopted. The shared encoder is a 12-layer Conformer~\cite{Conformer} of dimension 512. The label decoder for the neural Transducer is a stateless decoder~\cite{Stateless} with a dimension of 512. A convolution subsampling module of stride 4 is placed before the encoder to reduce the frame rate to 25 Hz. The model has 78.9M parameters in total.

\noindent\textbf{Training} Pruned RNN-T loss~\cite{PrunedRNN-T} interpolated with CTC loss~\cite{CTC} by factor 0.2 is used as the training objective function. Considering that the blank symbol prediction of the CTC branch is not accurate in the early stage, the frame skipping is applied after 4000 steps. A series of experiments are designed to explore the effect of hyperparameters ($\beta, \beta', \lambda, K$) in the following subsection. All models are trained with 4 NVIDIA V100 GPUs.

\noindent\textbf{Evaluation} The performances are evaluated on the LibriSpeech test-clean and test-other sets. Apart from Word Error Rate (WER), the frame reduction ratio (defined as $\frac{|\boldsymbol{f}_{skipped}|}{T}$) and real-time factor (RTF) are also measured as metrics for inference speed. In addition, to assess the ability of language model integration for the frame-skipped neural Transducer, WERs with external language models (LM) are also reported. Shallow fusion~\cite{ShallowFusion, ShallowFusion2, ShallowFusion3} and LODR~\cite{LODR} are investigated for LM fusion respectively. The external LM consisting of three LSTM layers~\cite{LSTM} is trained on LibriSpeech LM corpus and the low-order N-gram in LODR is a bi-gram LM trained on 960-hour transcription of LibriSpeech. The scales of the LSTM LM and the bi-gram LM are tuned using grid search on the dev set.

\subsection{Experimental Results}

\begin{table}[h]
\vspace{-0.5em}
  \centering
  \caption{Comparisons of WER(\%), Frame Reduction Ratio(\%), and RTF in neural Transducer using different strategies with different $\beta/\lambda/K$.}
  \label{tab:performance}
  \scalebox{0.88}{
  \setlength{\tabcolsep}{5pt}
  \renewcommand{\arraystretch}{1.2}
  \resizebox{\linewidth}{!}{
      \begin{tabular}{l|cc|c|c}
          \toprule
          \multirow{2}{*}{Method} &
          \multicolumn{2}{c|}{WER$\downarrow$} & 
          Frame Reduction &
          \multirow{2}{*}{RTF$\downarrow$} \\ 
          & clean & other & Ratio$\uparrow$ \\
          \midrule
          Baseline                      & 2.45 & 5.93 & 0.00     & 0.0106 \\
          \midrule
          Threshold & & & \\
          \hspace{2em} $\beta = 0.90$   & 2.47 & 6.07 & 65.66 & 0.0038 \\
          \hspace{2em} $\beta = 0.85$   & 2.54 & 5.95 & 66.64 & 0.0036 \\
          \hspace{2em} $\beta = 0.80$   & 2.59 & 6.11 & 68.34 & 0.0035 \\
          \midrule
          Soft Restriction & & & \\
          \hspace{2em} $\lambda = 0.03$ & 2.48 & 5.91 & 74.92 & 0.0026 \\ 
          \hspace{2em} $\lambda = 0.04$ & \textbf{2.44} & \textbf{5.88} & \textbf{75.44} & \textbf{0.0026} \\ 
          \hspace{2em} $\lambda = 0.05$ & 2.51 & 6.00 & 75.78 & 0.0024 \\ 
          \hspace{2em} $\lambda = 5.00$ & 2.90 & 6.89 & 78.25 & 0.0023 \\ 
          \midrule
          Hard Restriction & & & \\
          \hspace{2em} $K = 2$          & 2.49 & 5.92 & 72.35 & 0.0031 \\ 
          \hspace{2em} $K = 1$          & 2.92 & 6.96 & 78.23 & 0.0023 \\ 
          \bottomrule
      \end{tabular}}
  }
\vspace{-1em}
\end{table}
The results of applying different strategies to regularize blank are shown in \tbl{performance}. The baseline model is a neural Transducer\&CTC co-training system without frame skipping, where the CTC loss is used as an auxiliary loss with a factor of 0.2. As a comparison with existing frame skipping methods without modifying CTC topology, three models with different thresholds used in training and inference for frame skipping are also listed, referred as \textit{Threshold}. The average frame reduction ratios over the test sets are also shown in \tbl{performance}. As a reference number, we calculate the maximum possible reduction ratio $\gamma_{max} = 1 - \frac{S}{T} = 78.61\%,$ which is defined as the 1 minus the length ratio between output tokens and input frames. 

The following observations can be made:
\begin{itemize}
    \item Larger frame reduction ratios can be achieved after applying the soft or hard restrictions on the CTC topology compared to existing methods, indicating that a larger proportion of frames are recognized as blank frames by the CTC head; 
    \item Trade-offs between WERs and RTF can be tuned by adjusting $\lambda$ or $K$. A larger penalty $\lambda$ or a smaller $K$ encourages the shared encoder to discriminate between blank and non-blank frames, leading to more confident predictions for blank frames. By further increasing $\lambda$ to 5 or reducing $K$ to 1, the frame reduction ratio approaches $\gamma_{max}$ while maintaining reasonable accuracy;
    \item The proposed blank regularization method achieves 4 times speedup while yielding even slightly lower WERs than the baseline model without blank skipping. This indicates that the consecutively repeated non-blank frames contribute little to the decoding results with proper regularization during training, and discarding them during inference does not affect WERs.
\end{itemize}

\begin{figure}
    \caption{Aggregated WER(\%) versus Frame Reduction Ratio (\%) curve. Aggregated WER is the summation of WER on test-clean and test-other.}
    \label{fig:RP_tradeoff}
    \centering
    \begin{tikzpicture}[scale = 0.44]
  
    \definecolor{red1}{RGB}{154,0,0}
    \definecolor{red2}{RGB}{255,165,165}
    \definecolor{red3}{RGB}{205,0,0}
    \definecolor{red4}{RGB}{255,0,0}
    
    \definecolor{blue1}{RGB}{0,115,140}
    \definecolor{blue2}{RGB}{0,209,255}
    \definecolor{blue3}{RGB}{152,236,255}
    
    \definecolor{green1}{RGB}{0,111,44}
    \definecolor{green2}{RGB}{20,215,98}
    \definecolor{green3}{RGB}{192,247,214}
    
    \begin{axis}[
        legend style={
            nodes={scale=1.6, transform shape},
            at={(0.36, 0.98)}
        },
        width=\textwidth,
        tick align=inside,
        tick pos=left,
        xlabel=Frame Reduction Ratio (\%),
        ylabel=Aggregated WER (\%),
        xmin=58.5, xmax=79.5,
        ymin=8.20, ymax=10.00,
        xtick={59, 60, ..., 79},
        ytick={8.20, 8.30, ...,  10.10},
        label style={font=\Large},
        legend cell align={left},
    ]
        \draw[help lines, gray, opacity=.3] (58.5,8.20) grid (79.5,10.00);
    
        \addplot[ultra thick, mark=*, blue3]
        plot coordinates {
            (66.73325, 8.58)   
            (66.27035, 8.57)   
            (65.65960, 8.54)   
            (64.62653, 8.51)   
            (62.33147, 8.49)   
            (59.04870, 8.47)   
        };
        \addlegendentry{Threshold $\beta=0.90$}
        
        \addplot[ultra thick, mark=*, blue2]
        plot coordinates {
            (67.07390, 8.52)   
            (66.63868, 8.49)   
            (66.05883, 8.45)   
            (65.10254, 8.43)   
            (62.93088, 8.40)   
            (59.51931, 8.36)   
        };
        \addlegendentry{Threshold $\beta=0.85$}
        
        \addplot[ultra thick, mark=*, blue1]
        plot coordinates {
            (68.33661, 8.70)   
            (67.91106, 8.66)   
            (67.35993, 8.62)   
            (66.38959, 8.58)   
            (64.12835, 8.55)   
            (60.54135, 8.55)   
        };
        \addlegendentry{Threshold $\beta=0.80$}
        
        \addplot[ultra thick, mark=*, red1]
        plot coordinates {
            (78.32860, 9.88)   
            (78.25265, 9.79)   
            (78.13341, 9.74)   
            (77.90440, 9.66)   
            (77.17941, 9.53 )  
            (75.12709, 9.48 )  
        };
        \addlegendentry{Soft $\lambda=5.00$}
        
        \addplot[ultra thick, mark=*, red3]
        plot coordinates {
            (76.55638, 8.62)   
            (76.39750, 8.55)   
            (76.17245, 8.54)   
            (75.78175, 8.51)   
            (74.76605, 8.44)   
            (72.74183, 8.39)   
        };
        \addlegendentry{Soft $\lambda=0.05$}
        
        \addplot[ultra thick, mark=*, red4]
        plot coordinates {
            (76.24153, 8.46)   
            (76.07912, 8.43)   
            (75.84678, 8.40)   
            (75.44277, 8.32)   
            (74.37244, 8.26)   
            (72.28371, 8.24)   
        };
        \addlegendentry{Soft $\lambda=0.04$}
        
        \addplot[ultra thick, mark=*, red2]
        plot coordinates {
            (75.80693, 8.50)   
            (75.61736, 8.47)   
            (75.36702, 8.43)   
            (74.91681, 8.39)   
            (73.76294, 8.35)   
            (71.53770, 8.33)   
        };
        \addlegendentry{Soft $\lambda=0.03$}
        
        \addplot[ultra thick, mark=*, green2]
        plot coordinates {
            (73.67450, 8.49)   
            (73.38608, 8.47)   
            (73.00028, 8.46)   
            (72.35457, 8.41)   
            (70.81853, 8.34)   
            (68.32256, 8.30)   
        };
        \addlegendentry{Hard $K=2$}
        
        \addplot[ultra thick, mark=*, green1]
        plot coordinates {
            (78.31039, 9.95 )   
            (78.23277, 9.88 )   
            (78.11988, 9.78 )   
            (77.90680, 9.71 )   
            (77.20885, 9.59 )   
            (75.20232, 9.50 )   
        };
        \addlegendentry{Hard $K=1$}
        
        \addplot[ultra thick, color=black, loosely dashed]  
        table[row sep=crcr]{%
            0    8.38\\
            100  8.38\\
        };
        \addlegendentry{Baseline}

        \addplot[ultra thick, color=black, densely dashed]  
        table[row sep=crcr]{%
            78.61025   8\\
            78.61025   11\\
        };
        \addlegendentry{$\gamma_{max}$}
    \end{axis}
\end{tikzpicture}
\end{figure}

\figdot{RP_tradeoff} visualizes the relationship between frame reduction ratio and aggregated WER, the summation of WER on test-clean and test-other. The aggregated WER of the baseline model (horizontal loosely dashed line) and $\gamma_{max}$ (vertical densely dashed line) are also plotted for reference. Data points for each configuration are collected by varying the decoding blank threshold $\beta'$ from $\left(0.8,0.85,0.9,0.95, 0.99,0.999\right)$. As can be seen, the trade-off between aggregated WER and frame reduction ratio can be achieved by tuning $\beta'$ during decoding. A smaller $\beta'$ results in a larger proportion of blank frames while having higher WERs. Our proposed method achieves a much larger frame reduction ratio than the existing methods (Threshold-$x$) without sacrificing accuracy. By tuning $\lambda$ or $K$, the neural Transducer with blank-regularized CTC even outperforms the baseline without blank skipping while achieving over 75\% frame reduction ratio.

\begin{table}[h]
\vspace{-0.5em}
  \centering
  \caption{WERs(\%) and Relative WER Improvement from LM integration.}
  \label{tab:LM_gain}
  \setlength{\tabcolsep}{3pt}
  \renewcommand{\arraystretch}{1.2}
  \scalebox{0.88}{
  \resizebox{1\linewidth}{!}{
      \begin{tabular}{l|cc|c|cc|c}
          \toprule
          \multirow{2}{*}{Method} &
          \multicolumn{2}{c|}{Shallow Fusion$\downarrow$} &
          {Rel.} &
          \multicolumn{2}{c|}{LODR$\downarrow$} &
          {Rel.} \\
          & clean & other & {Imprv.$\uparrow$} & clean & other & {Imprv.$\uparrow$}\\
          \midrule
          Baseline                      & 2.24 & 5.35 & 9.43  & 2.16 & 5.09 & 13.48 \\
          \midrule
          Threshold & & & & & & \\
          \hspace{2em} $\beta = 0.85$   & 2.23 & 5.23 & 12.13 & 2.12 & 5.05 & 15.55 \\
          \midrule
          Soft Restriction & & & & & & \\
          \hspace{2em} $\lambda = 0.04$ & 2.19 & 5.17 & 11.54 & 2.09 & 4.94 & 15.50 \\
          \hspace{2em} $\lambda = 5.00$ & 2.55 & 6.16 & 11.03 & 2.44 & 5.87 & 15.12 \\
          \midrule
          Hard Restriction & & & & & & \\
          \hspace{2em} $K = 2$          & 2.19 & 5.21 & 12.01 & 2.07 & 5.01 & 15.81 \\
          \hspace{2em} $K = 1$          & 2.55 & 6.10 & 12.45 & 2.47 & 5.78 & 16.50 \\
          \bottomrule
      \end{tabular}}
  }
\vspace{-1em}
\end{table}
The WERs of decoding with external LMs are shown in \tbl{LM_gain}. The relative improvements of LM integration are calculated against the WERs in \tbl{performance}. Compared to the baseline model, the blank-regularized models achieve larger relative WER reduction with LM integration, suggesting that discarding more blank frames enables better fusion with external LMs.
\vspace{-0.25em}
\section{Conclusions}
Inspired by the observation that blank symbols in CTC and neural Transducer play similar roles, this paper proposed two novel blank-regularization methods to further boost the proportion of predicted blank symbols for the co-trained CTC model.
By discarding blank frames before going through the joint network, the blank-regularized CTC can accelerate the inference of the neural Transducer.
One interesting observation is that the frame reduction ratio of the neural Transducer can approach the theoretical boundary. 
Experiments show that the neural Transducer with guidance from the blank-regularized CTC achieves 4 times speedup during inference without sacrificing performance.
Our approaches achieve better trade-offs between WER and inference real-time factors than existing methods. Additionally, a further gain can be observed when decoding with external language models.


\newpage

\bibliographystyle{IEEEtran}
\bibliography{mybib}

\end{document}